\documentclass[aps,twocolumn,prl,superscriptaddress]{revtex4}
\usepackage{amsfonts}
\usepackage[final,hiresbb]{graphicx}
\usepackage{amsmath}
\usepackage{amssymb}
\usepackage{amstext}

\usepackage{color}
\usepackage{braket}

\definecolor{Green}{RGB}{0,204,102}
\definecolor{Purple}{RGB}{102,0,255}
\definecolor{Blue}{RGB}{51,153,255}
\definecolor{Red}{RGB}{151,010,010}

\begin{document}

\title{Control of Exciton Transport using Quantum Interference}

\author{Mark T. Lusk}
\affiliation{Department of Physics, Colorado School of Mines, Golden, CO 80401, USA}
\author{Charles Stafford}
\affiliation{Department of Physics, University of Arizona, Tuscon, AZ 85721, USA}
\author{Jeramy D. Zimmerman}
\affiliation{Department of Physics, Colorado School of Mines, Golden, CO 80401, USA}
\author{Lincoln D. Carr}
\affiliation{Department of Physics, Colorado School of Mines, Golden, CO 80401, USA}

\keywords{quantum interference, exciton, transistor, gating, transport, Wannier-Mott, Frenkel, Green function methods, mesoscopic, quantum control}

\begin{abstract}
It is shown that quantum interference can be employed to create an exciton transistor. An applied potential gates the quasi-particle motion and also discriminates between quasi-particles of differing binding energy. When implemented within nanoscale assemblies, such control elements could mediate the flow of energy and information. Quantum interference can also be used to dissociate excitons as an alternative to using heterojunctions. A finite molecular setting is employed to exhibit the underlying discrete, two-particle, mesoscopic analog to Fano anti-resonance. Selected entanglement measures are shown to distinguish regimes of behavior which cannot be resolved from population dynamics alone. 
\end{abstract} 

\maketitle

Within the quantum regime, the transport of charge can be gated by using an external field to control interferences inherent in its wave-like motion~\cite{Stafford}. Quantum interference can also be used to create transistors for quasi-particles such as electron Cooper pairs~\cite{Caviglia}, weakly to strongly interacting paired ultra cold fermions~\cite{Stadler, CarrLusk}, and spintronics~\cite{Wolf}. In this Letter, we demonstrate that such gating can be adapted to excitons.  These charge-neutral quasi-particles are typified by random walks~\cite{Fenna2, Lin}, but can also move coherently~\cite{Jang}. This allows the flow of information and energy to be considered within a circuit setting and is the first step towards establishing excitonic quantum control. It also has immediate implications for manipulating the dynamics of excitonic Bose-Einstein condensates as well as in quantum computing~\cite{Rivas}, artificial materials that incorporate polaritonic microcavities~\cite{Savvidis}, and transparent meta-materials~\cite{Panahpour}.  

Long-lived exciton motion has been observed in a number of inorganic, solid state systems such as ZnO, $\rm{Cu}_2{\rm O}$, inorganic quantum well structures~\cite{Gartner} and transition metal monolayers~\cite{Hanbicki}.  It is also central to all photosynthetic processes, where suggestions of quantum coherence~\cite{Lee}  led to the creation of carbon-based materials that support long-lived coherent superpositions of excitons~\cite{Yuen, Collini}. Whether by natural selection or by engineering design, though, the only way to guide exciton transport is by creating an energy landscape in which the quasiparticles travel downhill~\cite{Griffith}. Exciton control via lattice strain gradients~\cite{Fu} amounts to a continuum version of this inefficient and imprecise energy cascade concept. In the related field of optoelectronic transistors~\cite{Grosso, High}, stationary excitons mediate optical connectivity but the motion of the excitons themselves is not controlled.

\begin{figure}[t]\begin{center}
\includegraphics[width=0.5\textwidth]{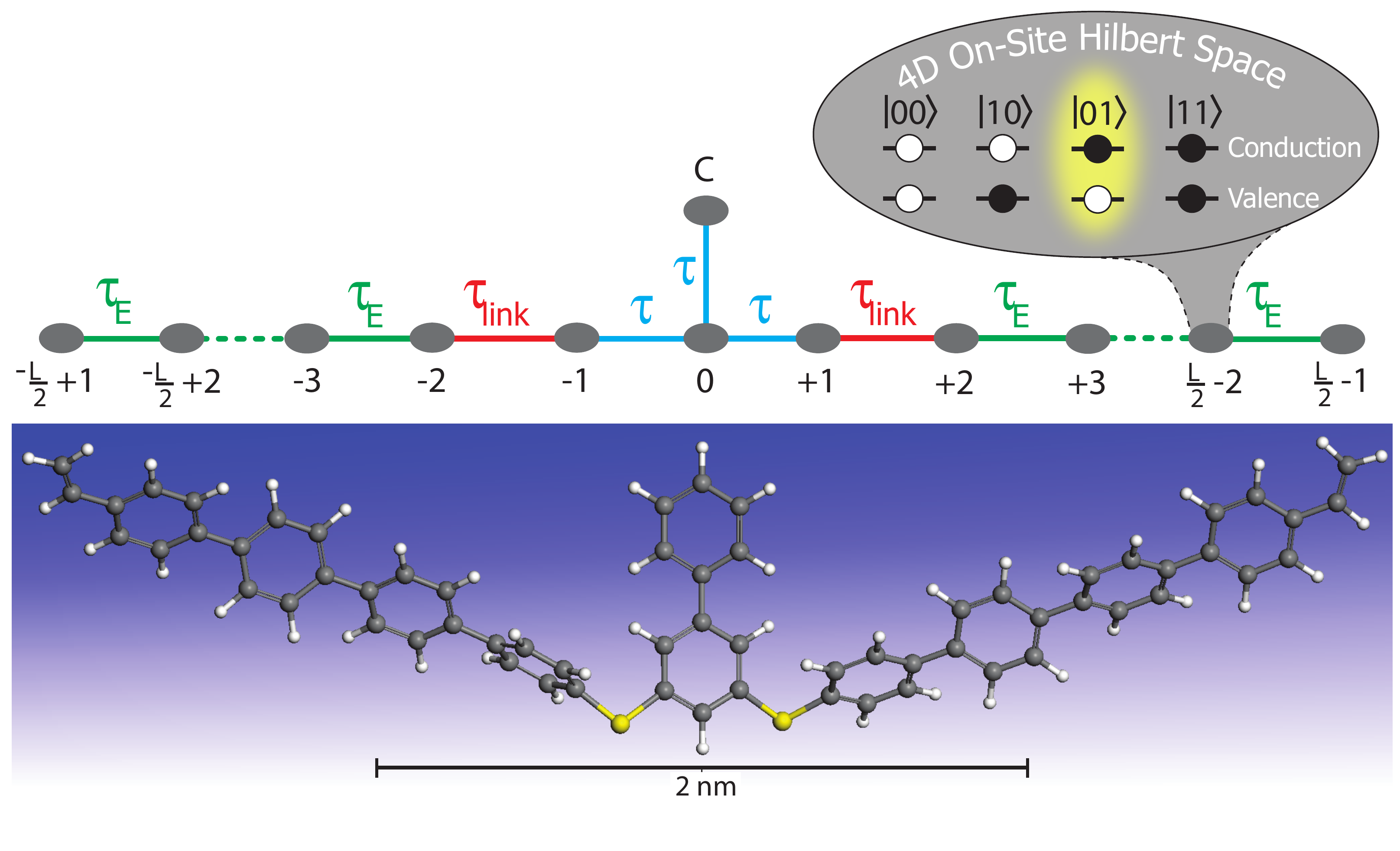}
\caption{
{\it Sketch of Exciton Transport}. Top: gating substructure (blue) composed of control (C) with potential energy adjusted via an external field, base (0) and bridge sites ($\pm 1$) linked (red) to semiconducting electrodes (green). Bottom: an idealized implementation in which vinyl-capped p-phenylene electrodes sandwich a 1,3-benzenedithiol junction topped with a phenyl control site. Zoomed inset: Each site supports 4 occupation states in the band structure (white = no electron, black = electron) with the exciton state highlighted.}
\label{Geometry}%
\end{center}
\end{figure}

We use a mesoscopic setting, shown in Fig.~\ref{Geometry}, to explore the transit of a single electron-hole superposition through a junction that generates quantum interference. In this regime, excitons cannot be idealized as being comprised of a continuous band of momentum states. Anti-resonance due to quantum interference is therefore distinct from that of Fano processes\cite{Bulka2001, Kobayashi2002, Stefanski2004, Rincon2009, Zhu2012} in which a discrete state interacts with a continuum, previously advanced as a means of gating the motion of charge~\cite{Stafford}. The degree of exciton transit is a strongly non-linear function of the adjustable potential energy of the control site which therefore acts as a gate.
%
%
 
We use Green function analysis, exact diagonalization, and scattering simulations to elucidate exciton transit within two regimes. For sufficiently strong Coulomb interaction, the exciton behaves as a single particle that can be blocked via anti-resonance using an external potential. Lower binding energies, though, allow for entirely different two-particle dynamics in which charge pairs can be selectively blocked based on their binding energy. These quasi-particles are, in addition, particularly susceptible to multiple forms of quantum control, as we will show.

The Hamiltonian of our quasi-1D system of L sites, as sketched in Fig.~\ref{Geometry}, takes the form
%
\begin{equation}
\hat{H}  = \hat{H}_\Delta  + \hat{H}_{e} +  \hat{H}_{\mathrm ex}  +  \hat{H}_U  + \hat{H}_V,
\end{equation}
\vskip -0.3 in
\begin{eqnarray}
\hat{H}_\Delta  &=& \textstyle{\sum}_{n,\nu}\Delta_{n}^{\nu}\hat{n}^{\nu}_{n}, \nonumber \\
\hat{H}_{e}   &=& \textstyle{\sum}_{<m,n>,\nu}\tau^{\nu}_{m n}\hat{c}^{\dagger \nu}_{m}\hat{c}_{n}^{\nu} + \rm{h.c.},    \nonumber \\
\hat{H}_{\mathrm ex}   &=& \textstyle{\sum}_{<m,n>,m\neq n}\chi_{m n}\hat{c}^{2\dagger}_{n}\hat{c}_{m}^{2}\hat{c}^{1\dagger}_{m}\hat{c}_{n}^{1}  + \rm{h.c.}, \\
\hat{H}_U   &=& \textstyle{\sum}_{n}U_{n}\hat{n}_{n}^1\hat{n}_{n}^2 + \rm{h.c.},  \nonumber \\
\hat{H}_V   &=& \textstyle{\sum}_{m\neq n,\nu, \mu} V_{mn}^{\mu\nu}\hat{n}_{m}^{\mu}\hat{n}_{n}^{\nu} + \rm{h.c.}\, \,.  \nonumber 
\label{Hcomponents}
\end{eqnarray}
%
$\hat{H}_\Delta$ is the band offset while $\hat{H}_e$ and $\hat{H}_{\mathrm ex}$ describe electron and exciton hopping. The former is a single particle operator that does not allow for charge carriers to change bands. However, $\hat{H}_{\mathrm ex}$ is a two-particle operator in which an excited electron is driven to its ground state and a neighboring ground state electron is raised to an excited state. This is, therefore, an explicit accounting of exciton motion rather than charge motion. The remaining two terms, $\hat{H}_U$ and $\hat{H}_V$, are on-site and potentially long-range Coulomb interactions. Roman subscripts $m$ and $n$ denote lattice sites, Greek superscripts $\mu$ and $\nu$ indicate electron band, $<\!\!\!m,n\!\!\!>$ means a sum over sites that are nearest neighbors, $\hat{c}_{n}^{\nu}$ is the electron annihilation operator for band $\nu$ of site $n$, $\hat{n}^{\nu}_{n} =\hat{c}^{\nu\dagger}_{n}\hat{c}_{n}^{\nu}$ is the electron number operator, and $[\hat{c}^{\mu}_{m}, \hat{c}^{\nu\dagger}_{n}]_+=\delta_{mn}\delta_{\mu \nu}$. Phonon and photon coupling are disregarded. 

Strong on-site Coulomb interactions result in Frenkel excitons~\cite{Frenkel} comprised of superpositions in which the electron and hole are on the same site, while weaker charge interactions result in Wannier-Mott excitons~\cite{Wannier}, superpositions in which the electron and hole may be substantially separated. Processes dominated by $\hat{H}_{e}$ will be referred to as {\it First-Order, Two-Particle} (2P) since exciton motion requires that the operator act separately on an electron and hole. This is typical of exciton dynamics in highly ordered, closely packed systems such as solid-state crystals. {\it Second-Order, One-Particle} (1P) processes are due to single quasi-particle hops of $\hat{H}_{\mathrm ex}$ associated with F$\ddot{\rm o}$rster resonant energy transfer, when Coulombic interactions dominate, and Dexter transfers, when exchange processes are most relevant. In general, both Frenkel and Wannier-Mott excitons can have 1P or 2P character or a combination of both. 

The nature of excitonic anti-resonance is distinctly different between the extreme cases of purely 1P and 2P processes, and analyzing them separately allows two antiresonant regimes to be identified. 1P transits correspond to standard anti-resonance and can be considered in isolation by restricting the  Hamiltonian to
\begin{equation}
\hat{H}_{\mathrm 1P} \! = \!\hat{H}_\Delta  +\hat{H}_{\mathrm ex}\,\, .
\label{H1}
\end{equation} 
We choose the free energy constant of $\hat{H}_{\mathrm 1P}$ such that all site energies $\Delta_n^\nu$ can be neglected except at the control site, where we take $\Delta_C^\nu = \Delta$.

In the absence of connecting electrodes in this 1P setting, the zero temperature retarded Green function response has a pole at $\omega = \Delta$. To be useful as a control element, though, the molecule needs to be encapsulated within left and right electrodes (Fig. \ref{Geometry}). If they are of infinite length, then the dynamics is that of a discrete structure interacting with a continuum--i.e. single particle Fano Anti-resonance. A straightforward Dyson series analysis~\cite{Cuevas} can then be used to generate the retarded Green function between sites $\mbox{-}1$ and $0$ as the links between the electrodes and molecule are activated. The results show that the presence and location of the antiresonant point is unaffected. 

\begin{figure}[hptb]\begin{center}
\includegraphics[width=0.48\textwidth]{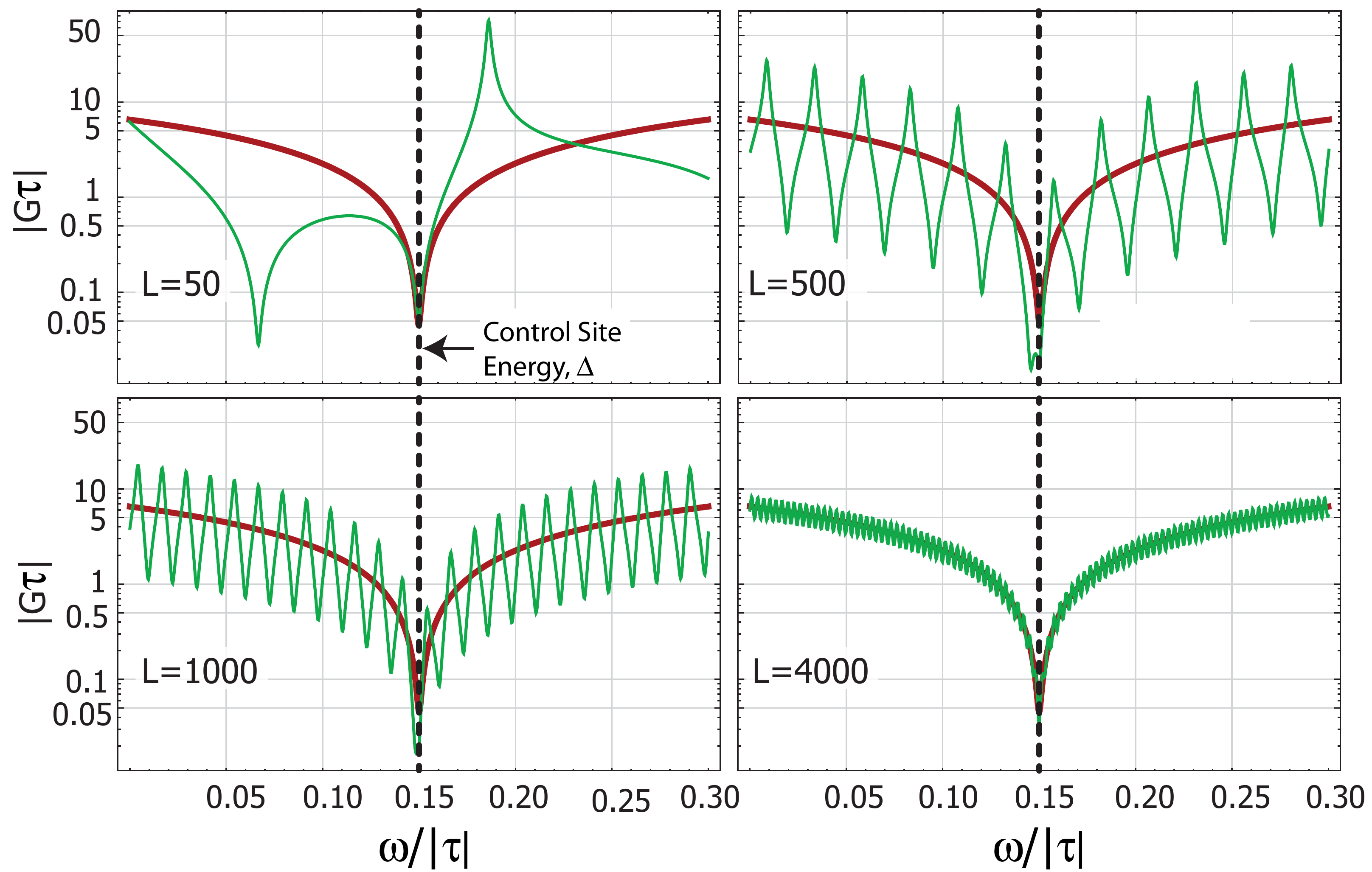}
\caption{
{\it 1P Green Function for Electrodes of Finite Length.} Control site energy $\Delta = 0.15/\tau$ and $\tau_{E}^{\nu}=\tau_{link}^{\nu}=\tau$. The red curve is an analytic Green function solution, $G_{\mbox{-}1,0}$, for the molecule placed between two semi-infinite electrodes. }
\label{G_2-Particle_Finite_Chains}
\end{center}\end{figure}

Finite length electrodes allow the mesoscopic 1P anti-resonance to be elucidated, and results obtained via exact diagonalization are shown in Fig. \ref{G_2-Particle_Finite_Chains}. Note the large number of singularities and roots as are observed for the Green function of a ring of identical sites~\cite{Hoffmann}. These are eventually obscured, making a band for any finite broadening, a standard result in condensed matter physics. The primary feature is an anti-resonance for excitons of an energy equal to that of the control site. The width of this region of transmission quenching is determined by the hopping parameter, $\tau$, which is why it is chosen as the characteristic energy in the plots; as $\tau$ decreases, the width of the anti-resonance region decreases.  

For relatively short electrodes, anti-resonance is obscured making it useful to supplement the Green function analysis with scattering simulations. This is taken up after a consideration of the other antiresonant extreme--that in which the electron and hole can move separately. 
%
%
%

Thus we turn to the 2P case, for which Coulomb interactions correlate the dynamics of upper and lower electrons but with richer physics than in the 1P setting. Such 2P processes are governed by the following reduced version of  the Hamiltonian:
%
\begin{equation}
\hat{H}_{\mathrm 2P} \! = \!\hat{H}_\Delta +\hat{H}_{e} + \hat{H}_U + \hat{H}_V \,\, .
\label{H2}
\end{equation}
%
The on-site Coulomb repulsion parameter, $U_n$, is nonzero for Frenkel excitons and is taken to be the same for each site. The non-local Coulomb repulsion parameter, $V_{mn}^{\mu\nu}$, describes Wannier-Mott excitons and is given a simple inverse distance dependence: $V_{mn}^{\mu\nu} = \beta\eta / (|m-n| + \eta)$. The total Coulomb energy is taken to be the binding energy of the exciton. In the absence of electrodes, a zero temperature retarded Green function response can be derived. Even with a neutral control site potential, this function exhibits anti-resonance at the binding energy of the exciton, surprising because it is elicited by a property of the electron-hole pair rather than that of the control site. We refer to this as Two-Particle Anti-resonance. Its Fano counterpart, the limiting case of infinite electrodes, has yet to be derived to the best of our knowledge. Unlike the 1P case, where the Hilbert space is of dimension $L$, the 2P case requires $L^2$, putting stricter numerical limits on exact diagonalization, and therefore severely restricting the electrode lengths that can be calculated. This motivates a re-analysis with exciton transit treated dynamically as a scattering event.

For both 1P and 2P extremes, a Gaussian wave packet is therefore sent through the gating assembly. The initial state can be a superposition of Frenkel excitons,
\begin{equation}
\ket{\psi(0)}  = \frac{1}{\pi^{\frac{1}{4}} \sigma^{\frac{1}{2}}}  \sum_m \mathrm{e}^{\imath k_0 m}\mathrm{e}^{\frac{-(m-m_0)^2}{2 \sigma^2}} \hat{c}^{2\dagger}_{m}\hat{c}_{m}^{1} \ket{\rm vac},
\label{initF}
\end{equation}
%
or of Wannier-Mott excitons,
\begin{equation}
\ket{\psi(0)} =\frac{1}{\pi^{\frac{1}{2}} \sigma} \sum
_{m,n} \mathrm{e}^{\imath k_0 (m+n)-\frac{(m-m_0)^2-(n-n_0)^2}{2 \sigma^2}}\hat{c}^{2\dagger}_{m}\hat{c}_{n}^{1} \ket{\rm vac} .
\label{initWM}
\end{equation}
Both types of initial conditions are created from a superposition of excitonic eigenstates of the relevant Hamiltonian. Here $\sigma$ is the exciton wave packet width, $m_0$ and $n_0$ denote wave packet centers, and wave number $k_0$ provides a right-directed kick. Note that the vacuum state, $\ket{\rm vac}$, is taken to be that for which all electrons reside in the valence band. 

A foundation for physically realizing such states exists within the quantum control community where laser pulse shaping has been explored as a means of exciting excitons on specific sites\cite{Hoyer:2014bf, Abramavicius:2008bc}. The associated optimization procedure can be extended to construct states with a prescribed envelope velocity as well. Laser pulses can therefore, in principle, be used to generate superpositions of excitonic eigenstates in a way that allows packet width and momentum to be tailored. Such quasiparticles are physically meaningful so long as phase coherence is maintained. These packets will spread as they move, so they need to be created relatively close to the control structure. This allows their scattering character to be quantified as is subsequently shown. 
\begin{figure}[hptb]\begin{center}
\includegraphics[width=0.38\textwidth]{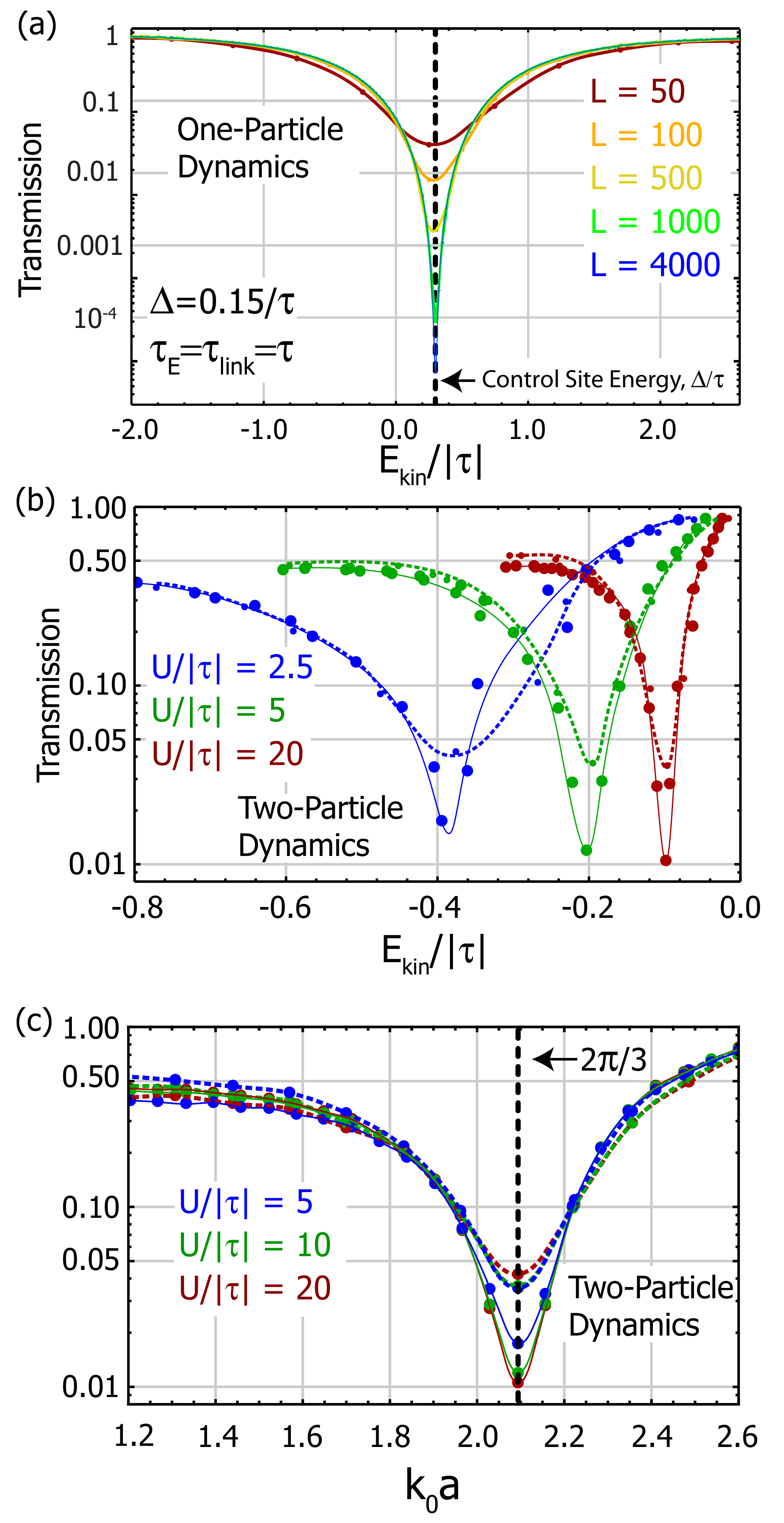}
\caption{
{\it Anti-resonances of Frenkel Excitons.} (a) 1P tuning.  (b) Kinetic energy required for 2P Anti-resonance increases with exciton binding energy, U. $L = 48$ (dashed) and $L = 99$ (solid). (c) Anti-resonance data of (b) can be aligned  by plotting transmission versus a normalized version of the kick parameter, $k_0$. Data is inherently discrete and curves are a guide to the eye.}
\label{Transmission_Panels}
\end{center}
\end{figure}

A range of kinetic energies can be chosen to construct transmission functions for each band based on the square of the respective projection amplitudes.  For instance, the top panel of Fig. \ref{Transmission_Panels} shows the transmission character that results when the 1P Hamiltonian of Eq.~\eqref{H1} is applied to scatter a Gaussian superposition of Frenkel excitons (Eq.~\eqref{initF}) off a biased control site. The transmission coefficient is taken to be the square of the ratio of the packet amplitude just before and after interaction with the gating assembly. Consistent with the Green function analysis, anti-resonance occurs when the kinetic energy of the exciton is equal to the biasing energy of the control site. This is clear in panel (a) and has been tested for a range of control site energies as well. While the associated wave packets are dispersive, the change in amplitude due to natural spreading is sufficiently slight that no correction was made to the transmission data.
 
The dynamics are richer when the 2P Hamiltonian of Eq.~\eqref{H2} is used to scatter a superposition of Frenkel excitons (Eq.~\eqref{initF}) with 2P Anti-resonance. As shown in Fig. \ref{Transmission_Panels}(b), the kinetic energy at which anti-resonance occurs depends on the binding energy of the exciton. The initial state has a total Coulomb energy of zero because Frenkel excitons generate no on-site repulsion. The subsequent motion of each electron/hole pair, though, amounts to the right transit of a conduction band electron combined with the left transit of a valence band electron---a non-radiative Dexter process~\cite{Dexter}. Intermediate states are therefore generated for which two sites are each partially occupied with two electrons, appearing and disappearing over one-half of a Rabi cycle. In order to move, then, the electron/hole pair must borrow kinetic energy so that it can temporarily delocalize. This is the case at each site and so is relevant at the control molecule as well. Anti-resonance occurs at a higher kinetic energy because some of it must be used to delocalize electron/hole pairs in order to carry out scattering. Fig. \ref{Transmission_Panels}(c) reveals, though, that the wave number for anti-resonance is the same for all six cases, $k_0 a = 2\pi/3$. This site-to-site phase shift, with $a$ the lattice spacing, is consistent with the two-step nature of exciton motion and is analogous to single-band dynamics with a two-site control chain extending up from the primary transmission channel. In general, single-band anti-resonance will occur for $k_0 a = \pi/(M+1)$ with an M-site control chain. Setting $M=2$ and noting that the phase shift must be doubled to account for the two-step exciton hops gives $k_0 a = 2\pi/3$. 

Although the present work focuses on anti-resonance, it is worth noting that the transmission coefficient is not necessarily equal to unity for frequencies far from anti-resonance. For instance, the 2P system of Fig. \ref{Transmission_Panels}(b), the transmission coefficient approaches $1.0$ as $k_0 a \rightarrow \pi$, but plateaus at approximately $0.5$ for values of $k_0 a$ below the anti-resonant point. These values include a correction made to remove the effect of natural packet spreading, so that is not the cause of an asymptote less than unity. Even for the 1P system of Fig. \ref{Transmission_Panels}(a), the transmission coefficient does not exceed $0.83$. It may be possible to mitigate such undesirable partial reflection of the wave packet with more sophisticated control structures. 

\begin{figure}[hptb]\begin{center}
\includegraphics[width=0.48\textwidth]{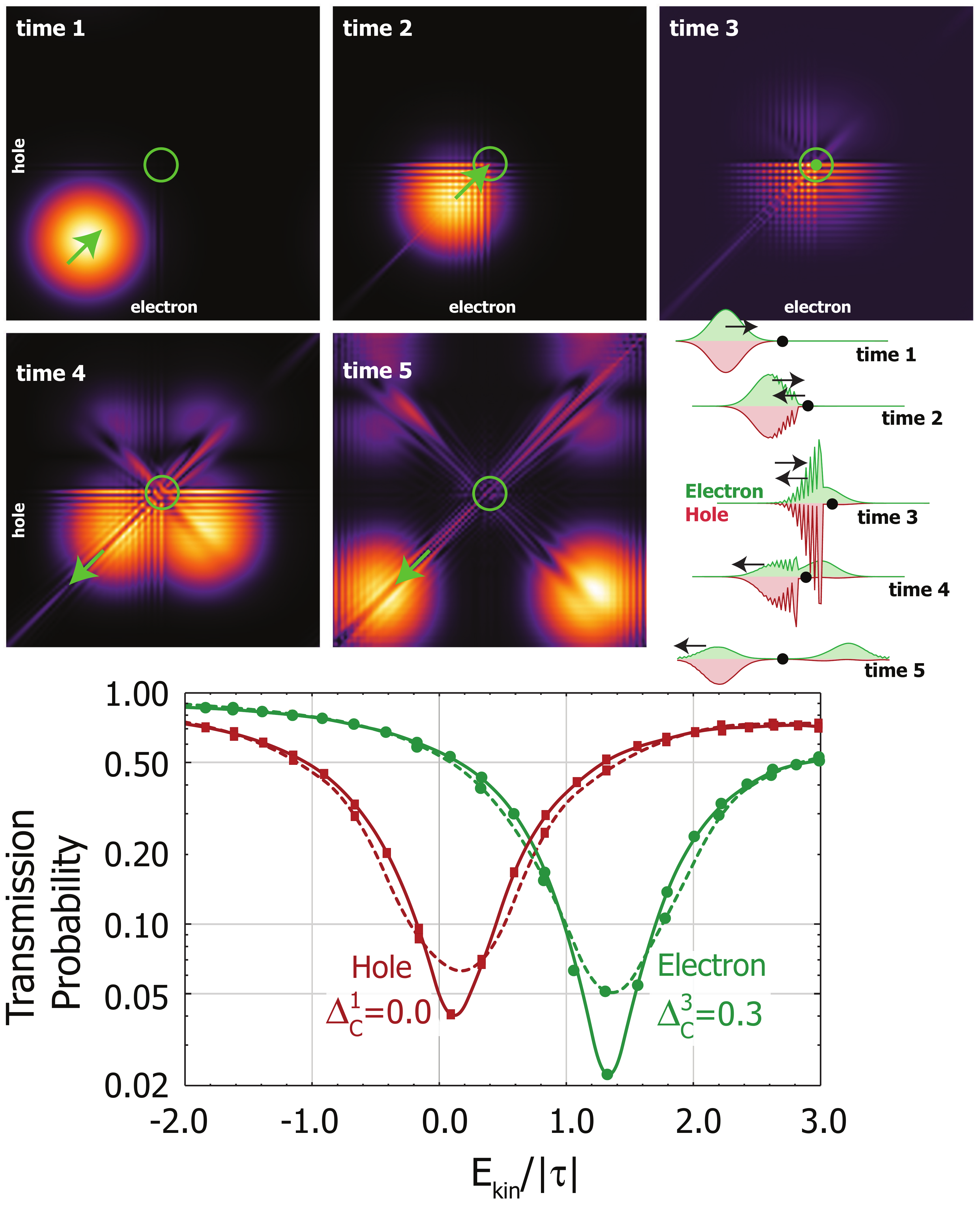}
\caption{%
{\it Exciton Dissociation.} Top: A symmetric Wannier-Mott exciton undergoes scattering  with different interference developed for hole and electron. Essentially all of the hole is reflected while most of the electron is transmitted. The lower right panel shows the same data as projections onto the electron and hole axes for each time slice. L = 100. Bottom: L = 50 (dashed) and L = 100 (solid). Data is inherently discrete and curves are a guide to the eye. $U_{n}+V_{mn}^{12} = 0.05|\tau|/(a|m-n|+0.5)$. The transmission ratio is 32:1.}
\label{exciton_dissociation}
\end{center}
\end{figure}

Control of the internal structure of excitons is also possible by constructing the gating assembly so that anti-resonance occurs at different energies for the two bands. In moderation, this will cause the exciton transmission with a partially dissociated electron-hole character. For sufficiently large energy mismatches, though, the exciton can be completely dissociated as shown in the 2P, Wannier-Mott scattering simulation of Fig. \ref{exciton_dissociation}. Using the same methodology as for the transmission data of Fig. \ref{Transmission_Panels}(b, c), a correction was made to remove the effect of packet spreading during transit through the control structure.

\begin{figure}[hptb]\begin{center}
\includegraphics[width=0.48\textwidth]{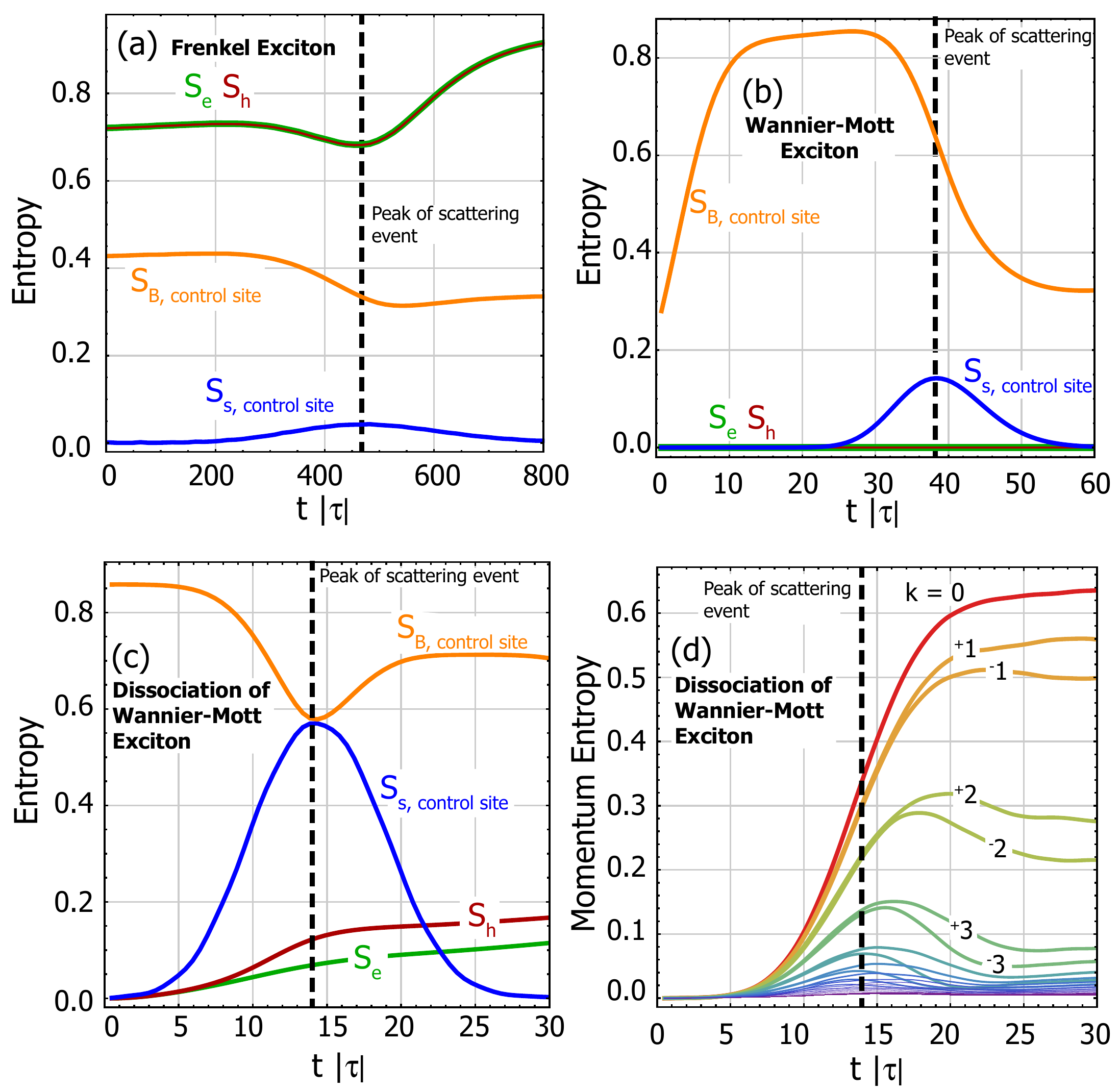}
\caption{%
{\it Entropies of Entanglement for 2P Dynamics.} Electron (e, green), hole (h, red), site (s, blue), bond (B, orange) and momentum (m, various colors) entropies. (a, b): anti-resonant scattering of Frenkel and  Wannier-Mott excitons; (c) delocalization and dissociation of Fig. \ref{exciton_dissociation}; (d) momentum entropies of Fig. \ref{exciton_dissociation}.}
\label{entanglement1}
\end{center}
\end{figure}

Entanglement measures provide additional insight into how the wave packet components interact during transit, and the entropy of entanglement~\cite{vonNeumann} is particularly useful in elucidating the nature of mixing of three reduced states with the rest of the system. The state operator is $\hat\rho (t) = \ket{\Psi (t)}\bra{\Psi (t)}$ and five partial traces were considered: electron states, hole states, states other than those associated with specified site or specified momentum and states associated with sites to the right of a specified site--i.e. for the bond entropy. The entropies are then given by 
\begin{equation}
S_n = -{\rm Tr} (\hat \rho_n {\rm log_j} \hat \rho_n)
\end{equation}
for the five partial traces (n = e, h, ${\mathrm s}$, ${\mathrm k}$,  $B$) with log bases chosen to be j = L, L, 4, 4 and $(B+1)^2$, respectively, so that maximal entropy is always unity, for convenience of comparison of relative values in Fig. \ref{entanglement1}.

Anti-resonance in the transit of a Frenkel exciton (Fig. \ref{entanglement1}a) is quantified as a dip in the electron and hole entropies and a peak in the site entropy of control site, C (see Fig. \ref{Geometry}). On a time scale shorter than depicted here, this site entropy exhibits a fine scale oscillation corresponding to the partial dissociation of the exciton required in order for it to move from site to site. The transit of a Wannier-Mott exciton (Fig. \ref{entanglement1}b) is distinctly different in that the electron and hole entropies are not affected by antiresonant scattering.  Exciton dissociation (Fig. \ref{entanglement1}c) can be identified with a local maximum (minimum) in the site (bond) entropy. Both hole and electron entropies accumulate in response to delocalization. Wannier-Mott excitons also exhibit nontrivial momentum entropies (Fig. \ref{entanglement1}d), where it is found that the entropy levels correlate with how close their momenta are to the mean value of the initial wave packet. These measures of entanglement offer precise metrics to characterize anti-resonance dynamics and a means of distinguishing Wannier-Mott from Frenkel excitons and 1P from 2P dynamics.

%



 
In conclusion, we have shown how to create an exciton transistor based on quantum interference. This could be used in excitonic logic circuits or to harvest hot excitons.  First- and second-order processes generate distinct anti-resonances that can be delineated using entanglement measures. Both 1P and 2P dynamics can be found in natural and artificial systems. Such exciton gating exists even for finite molecular chains---a discrete, quasi-particle, mesoscopic analog to Fano anti-resonance.  In addition to controlling exciton motion, it is also possible to dissociate them by exploiting quantum interference.  This could be used to produce electrical current without the energy loss that is inherent in heterojunction dissociation of excitons. Beyond the consideration of individual excitons, such control elements are expected to be relevant in developing techniques for mediating the transport of excitonic Bose-Einstein condensates, where exciton-exciton interactions and Bose coherence may lead to significant new quantum many-body features.

This material is based, in part, upon work supported by the National Science Foundation grant numbers PHY-1067973, PHY-1011156, and the Air Force Office of Scientific Research grant number FA9550-08-1-0069. CAS was supported by U.S. Department of Energy, Office of Science grant number DE-SC0006699.



\bibliographystyle{prsty}

\begin{thebibliography}{10}

\bibitem{Stafford}
C.~A. Stafford, D.~M. Cardamone, and S. Mazumdar, Nanotechnology {\bf 18},
  424014  (2007).

\bibitem{Caviglia}
A.~D. Caviglia, S. Gariglio, N. Reyren, D. Jaccard, T. Schneider, M. Gabay, S.
  Thiel, G. Hammerl, J. Mannhart, and J.~M. Triscone, Nature {\bf 456},  624
  (2008).

\bibitem{Stadler}
D. Stadler, S. Krinner, J. Meineke, J.-P. Brantut, and T. Esslinger, Nature
  {\bf 491},  736  (2012).

\bibitem{CarrLusk}
L.~D. Carr and M.~T. Lusk, Nature {\bf 491},  681  (2012).

\bibitem{Wolf}
S.~A. Wolf, D.~D. Awschalom, R.~A. Buhrman, J.~M. Daughton, S. von Molnár,
  M.~L. Roukes, A.~Y. Chtchelkanova, and D.~M. Treger, Science {\bf 294},  1488
   (2001).

\bibitem{Fenna2}
R.~E. Fenna and B.~W. Matthews, Nature {\bf 258},  573   (1975).

\bibitem{Lin}
Z. Lin, H. Li, A. Franceschetti, and M.~T. Lusk, ACS Nano {\bf 6},  4029
  (2012).

\bibitem{Jang}
S. Jang, Y.-C. Cheng, D.~R. Reichman, and J.~D. Eaves, J. Chem. Phys. {\bf
  129},    (2008).

\bibitem{Rivas}
D. Rivas, G. Muñoz-Matutano, J. Canet-Ferrer, R. García-Calzada, G. Trevisi,
  L. Seravalli, P. Frigeri, and J.~P. Martínez-Pastor, Nano Lett. {\bf 14},
  456  (2014).

\bibitem{Savvidis}
P.~G. Savvidis, L.~G. Connolly, M.~S. Skolnick, D.~G. Lidzey, and J.~J.
  Baumberg, Phys. Rev. B {\bf 74},  113312  (2006).

\bibitem{Panahpour}
A. Panahpour, Y. Silani, M. Farrokhian, A.~V. Lavrinenko, and H. Latifi, J.
  Opt. Soc. Am. B {\bf 29},  2297  (2012).

\bibitem{Gartner}
A. G${\rm \ddot a}$rtner, A.~W. Holleitner, J.~P. Kotthaus, and D. Schuh, Appl.
  Phys. Lett. {\bf 89},    (2006).

\bibitem{Hanbicki}
A. Hanbicki, M. Currie, G. Kioseoglou, A. Friedman, and B. Jonker, Solid State
  Commun. {\bf 203},  16   (2015).

\bibitem{Lee}
H. Lee, Y.-C. Cheng, and G.~R. Fleming, Science {\bf 316},  1462  (2007).

\bibitem{Yuen}
J. Yuen-Zhou, D.~H. Arias, D.~M. Eisele, C.~P. Steiner, J.~J. Krich, M.~G.
  Bawendi, K.~A. Nelson, and A. Aspuru-Guzik, ACS Nano {\bf 8},  5527  (2014).

\bibitem{Collini}
E. Collini and G.~D. Scholes, Science {\bf 323},  369  (2009).

\bibitem{Griffith}
O.~L. Griffith and S.~R. Forrest, Nano Lett. {\bf 14},  2353  (2014).

\bibitem{Fu}
X. Fu, C. Su, Q. Fu, X. Zhu, R. Zhu, C. Liu, Z. Liao, J. Xu, W. Guo, J. Feng,
  J. Li, and D. Yu, Adv. Mater. {\bf 26},  2572  (2014).

\bibitem{Grosso}
G. Grosso, J. Graves, A.~T. Hammack, A.~A. High, L.~V. Butov, HansonM, and
  A.~C. Gossard, Nat. Photonics {\bf 3},  577  (2009).

\bibitem{High}
A.~A. High, A.~T. Hammack, L.~V. Butov, M. Hanson, and A.~C. Gossard, Opt.
  Lett. {\bf 32},  2466  (2007).

\bibitem{Bulka2001}
B.~R. Bulka and P. Stefanski, Phys. Rev. Lett. {\bf 86},  5128  (2001).

\bibitem{Kobayashi2002}
K. Kobayashi, H. Aikawa, S. Katsumoto, and Y. Iye, Phys. Rev. Lett. {\bf 88},
  256806  (2002).

\bibitem{Stefanski2004}
P. Stefanski, A. Tagliacozzo, and B.~R. Bulka, Phys. Rev. Lett. {\bf 93},
  186805  (2004).

\bibitem{Rincon2009}
J. Rincon, K. Hallberg, A.~A. Aligia, and S. Ramasesha, Phys. Rev. Lett. {\bf
  103},  266807  (2009).

\bibitem{Zhu2012}
J.-X. Zhu, J.-P. Julien, Y. Dubi, and A.~V. Balatsky, Phys. Rev. Lett. {\bf
  108},  186401  (2012).

\bibitem{Frenkel}
J. Frenkel, Phys. Rev. {\bf 37},  17  (1931).

\bibitem{Wannier}
G.~H. Wannier, Phys. Rev. {\bf 52},  191  (1937).

\bibitem{Cuevas}
J. Cuevas, {\em Molecular electronics an introduction to theory and experiment}
  (World Scientific, Singapore Hackensack, NJ, 2010).

\bibitem{Hoffmann}
R. Hoffmann, {\em Solids and surfaces : a chemist's view of bonding in extended
  structures} (VCH Publishers, New York, NY, 1988).

\bibitem{Hoyer:2014bf}
S. Hoyer, F. Caruso, S. Montangero, M. Sarovar, T. Calarco, M.~B. Plenio, and
  K.~B. Whaley, New Journal of Physics {\bf 16},  045007  (2014).

\bibitem{Abramavicius:2008bc}
D. Abramavicius, D.~V. Voronine, and S. Mukamel, Proceedings of the National
  Academy of Sciences {\bf 105},  8525  (2008).

\bibitem{Dexter}
D.~L. Dexter, J. Chem. Phys. {\bf 21},  836  (1953).

\bibitem{vonNeumann}
J. von Neumann, {\em Mathematical Foundations of Quantum Mechanics}, {\em
  Princeton Landmarks in Mathematics and Physics Series} (Princeton University
  Press, Princeton, New Jersey, 1955), translated from the German edition by
  Robert T. Beyer. Original first edition published in German in 1932.

\end{thebibliography}

\end{document}